\documentclass[aps,prl,twocolumn,showpacs,superscriptaddress,superscriptaddress]{revtex4-1}  
\usepackage{graphicx,color}  
\usepackage{dcolumn}   
\usepackage{bm}        
\usepackage{amssymb}   

\usepackage{ulem}
\usepackage[section]{placeins}
\usepackage{float}
\usepackage{amsmath}
\usepackage{epstopdf}
 
\usepackage{natbib}

\begin{document}

\title{Phoretic Interactions Between Active Droplets}

\author{Pepijn G. Moerman}
\affiliation{Center for Soft Matter Research, Department of Physics, New York University, New York, NY 10003}
\affiliation{Debye Institute for Nanomaterials Science, Utrecht University, 3584 Utrecht, Netherlands}
\author{Henrique W. Moyses}
\affiliation{Center for Soft Matter Research, Department of Physics, New York University, New York, NY 10003}
\author{Ernest B. van der Wee}
\affiliation{Debye Institute for Nanomaterials Science, Utrecht University, 3584 Utrecht, Netherlands}
\author{David G. Grier}
\affiliation{Center for Soft Matter Research, Department of Physics, New York University, New York, NY 10003}
\author{Alfons van Blaaderen}
\affiliation{Debye Institute for Nanomaterials Science, Utrecht University, 3584 Utrecht, Netherlands}
\author{Willem K. Kegel}
\affiliation{Debye Institute for Nanomaterials Science, Utrecht University, 3584 Utrecht, Netherlands}
\author{Jan Groenewold}
\affiliation{Debye Institute for Nanomaterials Science, Utrecht University, 3584 Utrecht, Netherlands}
\affiliation{Academy of Advanced Optoelectronics, South China Normal University, Guangzhou, China}
\author{Jasna Brujic}
\affiliation{Center for Soft Matter Research, Department of Physics, New York University, New York, NY 10003}
\date{\today}

\begin{abstract}
Concentration gradients play a critical role in embryogenesis, bacterial locomotion, as well as the motility of active particles. Particles develop concentration profiles around them by dissolution, adsorption, or the reactivity of surface species. These gradients change the surface energy of the particles, driving both their self-propulsion and governing their interactions. Here we uncover a regime in which solute-gradients mediate interactions between slowly dissolving droplets without causing autophoresis. This decoupling allows us to directly measure the steady-state, repulsive force, which scales with interparticle distance as $F\sim {1/r^{2}}$. Our results show that the process is diffusion rather than reaction rate limited, and the theoretical model captures the dependence of the interactions on droplet size and solute concentration, using a single fit parameter, $l=16\pm 3$~nm, which corresponds to the lengthscale of a swollen micelle.  Our results shed light on the out-of-equilibrium behavior of particles with surface reactivity.
\end{abstract}

\maketitle

Concentration gradients develop around particles that locally alter the composition of their solvent.  This can occur if the particles dissolve in the solvent, if they adsorb other species in solution, or if their surfaces catalyze chemical reactions.  Examples include heterogeneous catalysts~\cite{Wang2013,Soto2014}, droplets undergoing Ostwald ripening, silica particles dissolving in a strong base, ion-exchange resin particles~\cite{Reinmüller2013} and microbes that are consuming nutrients or excreting signaling proteins~\cite{Kravchenko-Balasha2016,Kravchenko-Balasha2014}. These concentration profiles can affect the behavior of the dispersed particles if their surface tension couples to the solute concentration ~\cite{Soto2014,Abécassis2008,Cira2015,Toyota2009}. The most studied example is given by autophoretic swimmers, which form asymmetric concentration profiles and subsequently swim in response to the gradients~\cite{Elgeti2015,Abbott2016,Paxton2006,Sengupta2012}. As a result they move in a directional manner which has been shown to cause the formation of dynamic patterns~\cite{Wang2013,Palacci2012,Thutupalli2011,Buttinoni2013}. Currently there is considerable interest in self-propelled particles because they constitute model systems for studying collective behavior from a range of fields and disciplines. Examples include pattern formation~\cite{Sanchez2012}, dynamic clustering~\cite{Palacci2012,Buttinoni2013,Theurkauff2012} and anomalous density fluctuations~\cite{Schaller2013} with connections to glassy behavior and jamming~\cite{Ni2013}. Despite this interest, the propulsion mechanism of many important model systems is not well understood~\cite{Brown2014}, and their mutual interactions even less so. 

In the case of particles that are self-propelled by chemical gradients, it is clear that overlap of concentration profiles around two or more particles results in mutual interactions~\cite{Soto2014,Cira2015}. These concentration gradients evolve as the particles swim, which precludes a precise definition of the interparticle potential (and in particular, the motion of the particle is non-Markovian). To this end, we uncover the regime in which active droplets do not swim, but do exude concentration profiles. This system allows us to measure these gradient-mediated interactions between pairs of droplets in the absence of autophoresis. We then develop a theoretical model based on steady-state diffusion profiles to the functional form for the interaction and thus fit the data as a function of droplet size and solute concentration. This `static' case improves our understanding of the mechanism that leads to a threshold concentration above which droplets begin to swim. 

\begin{figure}[b!]
\includegraphics[scale=1]{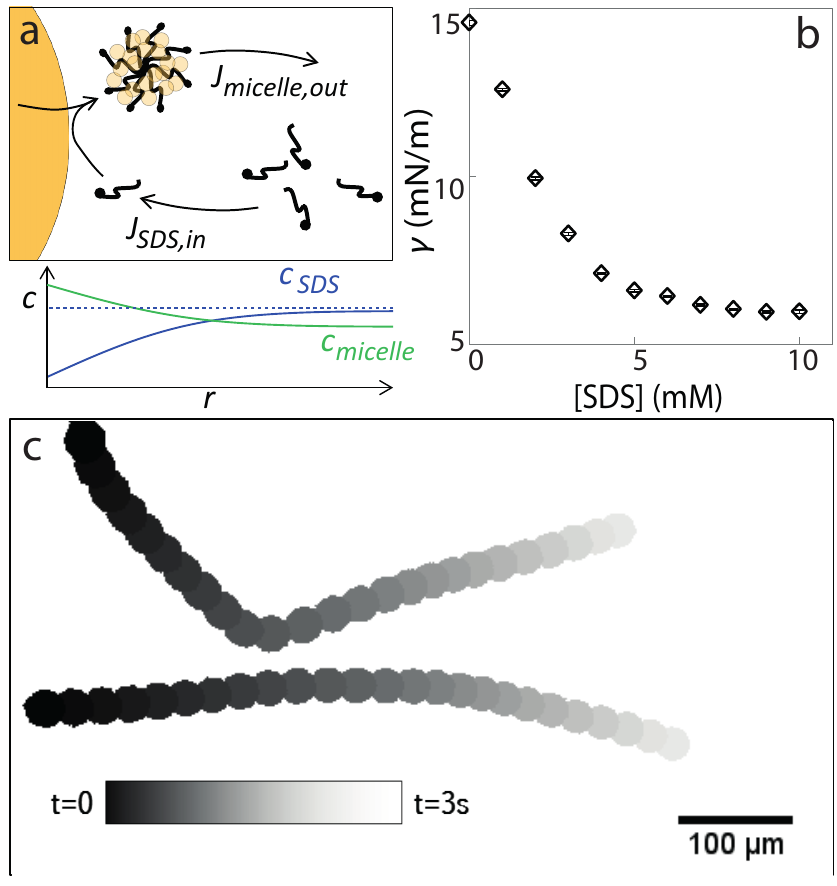}
\caption{\label{fig1} Droplet interactions due to surfactant gradients. a) DEP oil dissolves into swollen SDS micelles, giving rise to radial concentration gradients of SDS monomer (blue line) and micelles (green line) surrounding the droplet surface, compared to the bulk concentration (dashed line). b) Surface tension of DEP droplets in water decreases as a function of SDS concentration. c) Two oil droplets swimming in given initial directions repel one another as a result of their concentration gradients. Circles map their trajectories over time. }
\end{figure}

We employ a simple model system of droplets of di-ethyl phthalate (DEP) oil dispersed in an aqueous solution of the surfactant sodium dodecyl sulphate (SDS). The DEP droplets slowly dissolve in the medium, giving rise to local concentration gradients. DEP is only marginally soluble in water (0.2 mg/mL~\cite{SM}). Above a threshold SDS concentration of $4$~mM, surfactant molecules and DEP molecules from the droplet coassemble to form oily micelles, causing the droplets to shrink further at a rate that depends on the SDS concentration. This process, schematically depicted in Fig.~\ref{fig1}a, depletes the surfactant molecules near the surface and results in a radially symmetric concentration profile of SDS. Figure~\ref{fig1}b shows that the surface tension between water and DEP decreases with increasing SDS monomer concentration, as measured using the pendant drop method  ~\cite{Faour1996}. This coupling between the surface free energy of the particle and the surfactant monomer causes droplets to move towards higher SDS concentrations in the bulk. 

Initially, the dissolution leads to an isotropic concentration profile and no net force acts on the particle. Beyond a cutoff dissolution rate, however, the isotropic state becomes unstable and any fluctuation gives rise to self-sustained motion in a random direction~\cite{Izri2014,Michelin2013}. These ballistic trajectories repel one another, as shown in the example of two swimming DEP droplets in Fig.~\ref{fig1}c and in Supplementary Movie 1. To develop an understanding of the interparticle coupling, here we focus on the regime of SDS concentrations in which the droplets are surrounded by a symmetric concentration profile and do not swim. Experimentally, this regime exists between $4$~mM SDS, below which the droplets are insoluble, and $8$~mM SDS, above which the droplets swim. 

\begin{figure}
\includegraphics[scale=1]{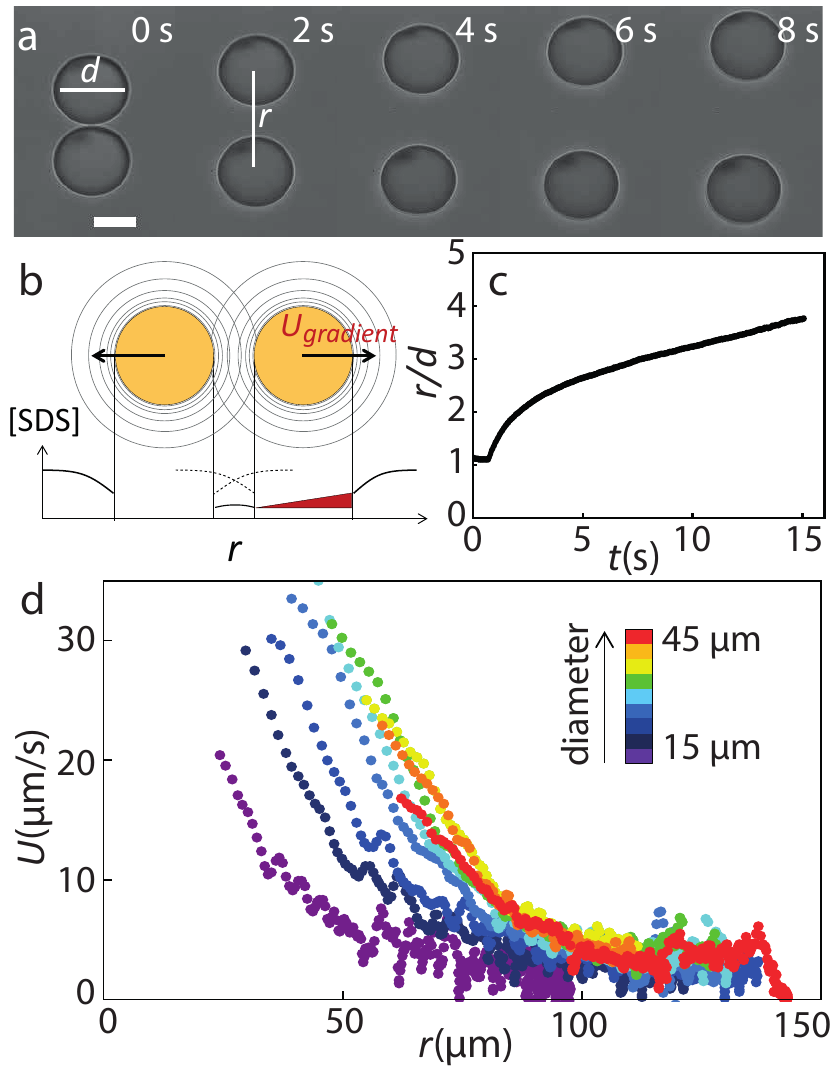}
\caption{\label{fig2} (Color online) Phoretic interactions. a) Frames of a movie showing two oil droplets moving away from each other due to solute-mediated repulsion after they have been brought into contact using optical tweezers. b) Schematic drawing showing how overlapping SDS concentration profiles lead to interaction. The red triangle represents the SDS gradient causing the motion. c) Droplet separation as a function of time. d) Velocity $U$ with which droplets move away from each other as function of distance. Colors represent droplet size.}
\end{figure}

In this regime we measured the interaction strength between dissolving droplets using blinking optical tweezers~\cite{Crocker1994,Krishnatreya2014}. Two holographically projected optical traps were used to bring two droplets close together and then released to allow the particles to move under the influence of the interaction force. Figure~\ref{fig2}a and Supplementary Movie 2 show a typical time sequence as the particles move apart during one such cycle. Figure~\ref{fig2}b gives a schematic overview of the overlapping concentration profiles that induce an effective interaction. We obtain an estimate for the interaction force by analyzing images of the particle motion.  The time trace of the center-to-center separation, $r(t)$, is plotted in  Fig.~\ref{fig2}c.  The derivative of this trajectory yields the relative separation speed,$U(r)$, examples of which are plotted in Fig.~\ref{fig2}d as a function of droplet size. The droplets range in diameter from $15-45$$\mu$m, and so exhibit no Brownian motion. They move with maximum speeds below $40$$\mu$m/s, and thus still have a low Reynolds number. Their relative speed is therefore directly proportional to their effective interaction force.

The larger the droplets, the stronger the repulsive interaction, as shown by the data in Fig.~\ref{fig2}d. In all cases, the range of the interaction exceeds $50$$\mu$m, which is longer than that expected for electrostatic interactions. The Debye-Huckel screening length is less than $10$~nm at the ionic strengths of our experiments. The typical velocity scale of $10$$\mu$m/s corresponds to forces on the order of $10$~pN. The fact that experiments performed at SDS concentrations below $5$~mM show no repulsion confirms that the repulsion is concomitant with the formation of DEP-swollen micelles of SDS. As the SDS concentration is increased, the DEP dissolves faster into the micelles, creating a steeper gradient, which results in an increased interaction strength~\cite{SM}. 

\begin{figure}
\includegraphics[scale=1]{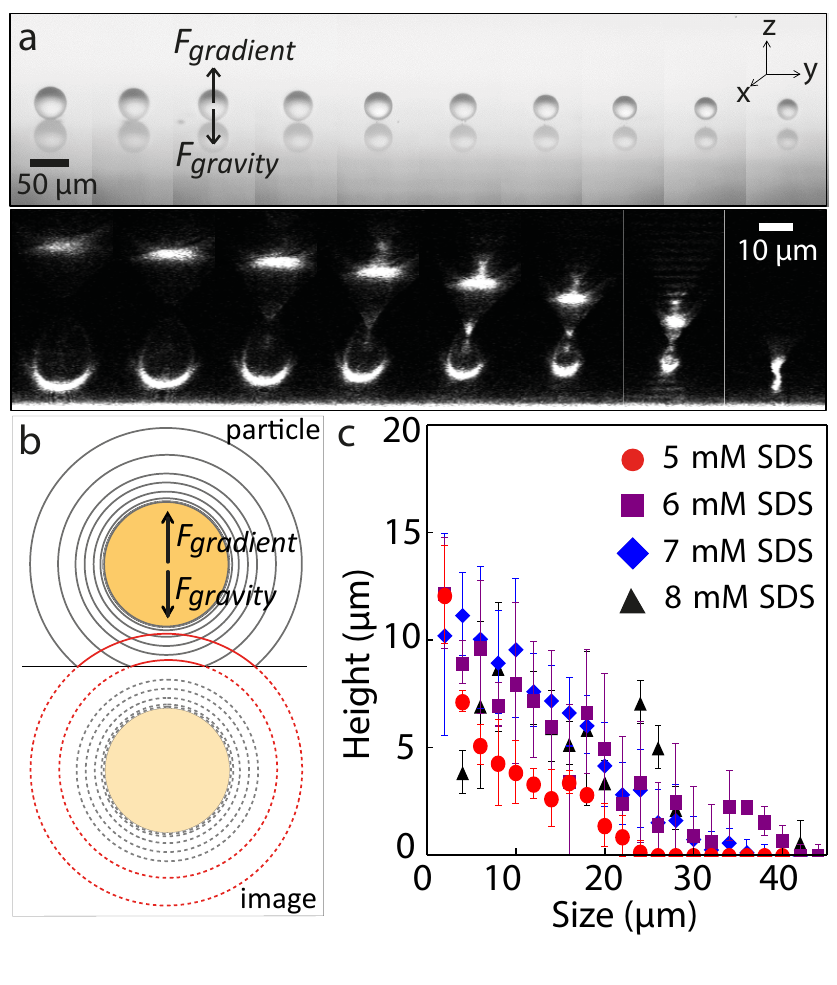}
\caption{\label{fig3} (Color online) Hovering due to solute-mediated interactions. a) Top panel shows a dissolving DEP droplet through a tilted bright-field microscope. The top feature is the real droplet and the lower feature the optical image on the glass slide. Bottom panel shows $xz$ projections of the same droplets imaged in reflection mode on a confocal microscope. b) Schematic drawing shows how a solvent cloud around a particle can lift it from the glass surface against gravity. c) Measured droplet hovering heights as a function of their size. Error bars come from repeated measurements.} 
\end{figure}

The force mediated by solvent gradients is of the same order of magnitude as the gravitational force acting on individual droplets. This allows for an alternative measurement of the strength of a solute-mediated interaction by balancing it with gravity. Figure~\ref{fig3}a shows images of dissolving droplets through a tilted microscope. The top feature is an image of the actual droplet and the bottom feature is an optical reflection in the glass slide. The droplet height is then half the distance between the droplet and its image. Figure~\ref{fig3}a shows that particles with a diameter over 30~$\mu$m make contact with the glass slide because their weight is larger than the gradient force, which repels the particles from the glass wall. For smaller particles, however, the two forces are comparable, resulting in an equilibrium hovering height above the glass cover slide at which the two forces balance. 

The lower panel in Fig.~\ref{fig3}a shows the same set of experiments performed using confocal microscopy in reflection mode. The elongated shape of the image is caused by internal reflection inside the droplet and the correct measure of the droplet size is the width of the bottom half-sphere~\cite{SM}. The schematic drawing in Fig.~\ref{fig3}b indicates how the same solute-mediated interaction that causes repulsion between two droplets is responsible for the hovering above a glass surface. The SDS concentration around the droplet is lower near the glass slide, because no influx of SDS molecules is possible through the glass. This decreased SDS concentration also occurs between two dissolving droplets, as shown in Fig.~\ref{fig2}b, so the two situations result in a similar solute-mediated force. Figure~\ref{fig3}c shows the equilibrium height of droplets of various SDS concentrations and various sizes, highlighting the trend that smaller active droplets at higher SDS concentrations hover at higher altitude.

Next, we present a theoretical model that predicts the functional form of the interaction and compare it with our measurements of the interaction strength between two droplets. The speed of a droplet $U(r)$ in the concentration profile of its neighbor is proportional to  the product of the concentration gradient $\nabla c$ and the particle mobility $M$. The mobility is given by ~\cite{Michelin2013}
\begin{equation}
\label{eq:bareM}
M = \frac{2 a K}{3(2\eta_o+3\eta_i)}
\end{equation} 
where $a$ is the droplet radius, $K$ is the slope of the surface tension versus SDS concentration graph in Fig.~\ref{fig1}b, and $\eta_o$ and $\eta_i$ are the viscosities of the continuous and dispersed phases, respectively. Assuming steady-state, and imposing the general boundary condition that the diffusive SDS flux to the surface must equal the rate at which SDS is consumed to form swollen micelles~\cite{SM}, the gradient is given by
\begin{equation}
\nabla c=  \frac{(c_{\infty}-c^*)}{1+l/a}\frac{a}{r^2} \label{eq1}
\end{equation}
where  $c_{\infty}$ is the bulk SDS concentration and $c^*$ is a threshold concentration, which is similar to the critical micelle concentration, but applies to DEP-swollen micelles of SDS. For these micelles, $c^*=4$~mM~\cite{SM}. The quantity $l=D/k$ is a length obtained by dividing the diffusion coefficient $D$ by the dissolution speed $k$, i.e. the speed at which oil moves across the droplet surface. 

Whether the process is limited by a reaction or by diffusion manifests itself in this length scale $l$. The limit $l/a\ll1$ corresponds to a constant surface concentration and thus to diffusion-limited dissolution. The opposite limit,  $l/a\gg1$, corresponds to a constant dissolution rate, so a reaction-limited system. Plotting the size of the DEP droplets as a function of time reveals a linear dependence of $a^2-a_0^2$, which is consistent with a diffusion limited process~\cite{SM}. By contrast, previous work assumes a reaction limited process~\cite{Izri2014,Michelin2013}, which gives rise to a linear relation of $a-a_0$ versus time.

Since the active droplet is also dissolving and creating its own gradient of the same solute, its speed in Eq.~\eqref{eq1} is somewhat modified. The surface reaction modifies the external gradient at the particle surface in a way that depends on the value of $l$. By solving the steady-state diffusion equation in the presence of an external gradient~\cite{SM} we find that the bare $\nabla c$ given in Eq.~\eqref{eq1} must be multiplied by the factor
\begin{equation}
\frac{\partial G_s}{\partial G_e}=\frac{3l/a}{1+2l/a}\approx\frac{3l}{a}\label{eq2}
\end{equation}
where $G_s$ is the surface gradient corresponding to the back-front asymmetry in the solute concentration on the droplet surface, and $G_e$ is the externally applied gradient. Note that in the case where $l/a\gg1$ the surface gradient is enhanced by the surface reaction, whereas in our case $l/a\ll1$ and the surface gradient is smaller than the external gradient, highlighting the importance of distinguishing between the two regimes. This effect modifies the concentration profile by a factor $3l/a$, which is independent of the droplet separation $r$.    

In addition, the fact that the particle is in motion at a given velocity further enhances the response to the external gradient. As an active particle moves, advection causes accumulation of the solute at the back of the particle, which leads to a back-front asymmetry in the solute concentration around the particle. This asymmetry directs the self-sustained motion at a high enough surfactant concentration~\cite{Michelin2013} and this coupling between flow and solute gradient gives rise to the swimming instability for isotropic particles. Even below the self-propulsion threshold for a single particle, the speed at which two particles move away from each other is enhanced by this effect. We consider the solute gradient around an active particle moving with a constant velocity and find that the external gradient increases linearly with the particle speed~\cite{SM}. Above a threshold speed, the motion becomes self-sustained. For a diffusion-limited process, the Peclet number of this transition is independent of droplet size and only scales with solute concentration. Indeed, unlike any other known swimmers, experimentally we show that droplets of all sizes swim above the SDS concentration of $c_{cr}=9$~mM. These results are the first to demonstrate a system in which active particles are not simply on or off, but only swim above a given concentration. 

In terms of the droplet-droplet repulsive interaction, this effect gives rise to a nonlinear correction factor to $U(r)$,
\begin{equation}
\label{eq:m2}
	\alpha = \frac{c_{cr}-c^*}{c_{cr}-c_{\infty}}.
\end{equation}

Interestingly, the correction due to the presence of the wall turns out to be negligible~\cite{SM}. On the one hand, the hydrodynamic drag on the particles slows them down, depending on the distance to the wall. Assuming the heights of a moving and a stationary droplet are comparable, we use the data in Fig.~\ref{fig3}c to calculate this correction due to the drag. On the other, the depletion of surfactant near the wall, responsible for the hovering effect shown in Fig.~\ref{fig3}, enhances the repulsion and speeds up the droplets. These two effects are of comparable magnitude and therefore cancel each other out. 

\begin{figure}
\includegraphics[scale=0.7]{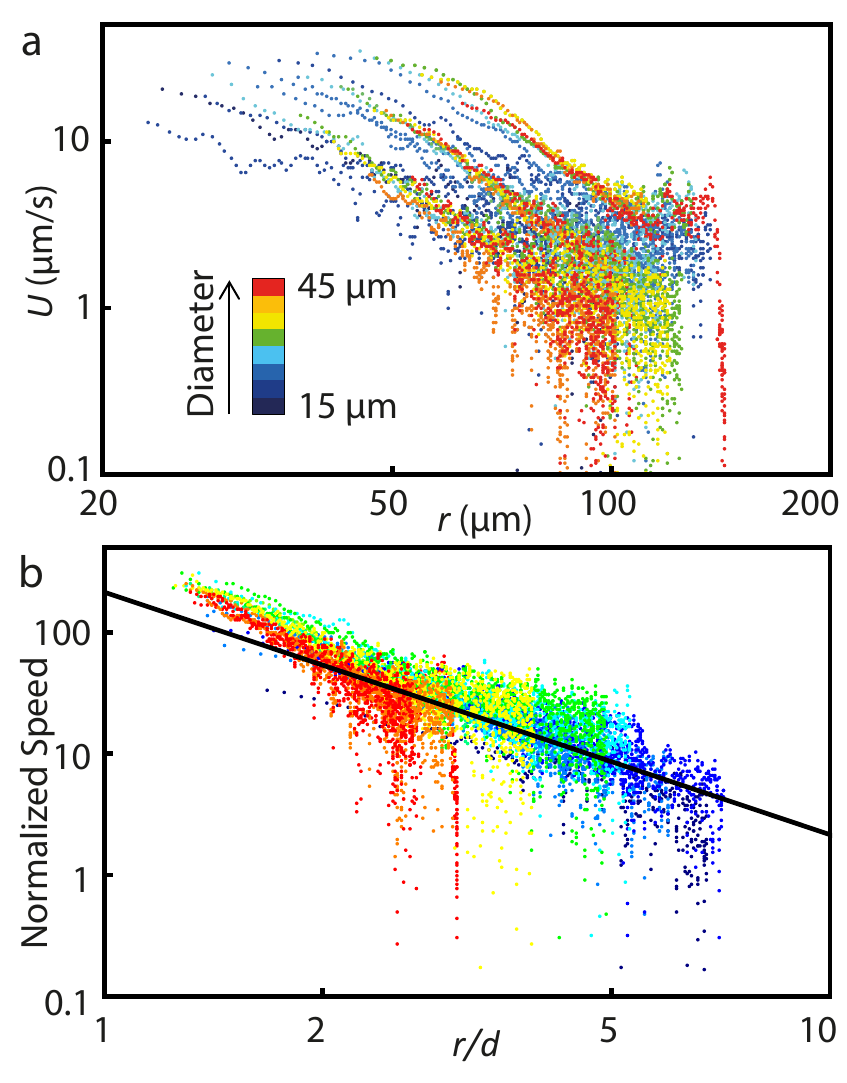}
\caption{\label{fig4} (Color online) Rescaling of active interactions a) Log-log representation of the speed with which droplets move away from each other as a function of interparticle distance. Color represents the initial droplet size. The lower of the three clusters of points is the set of measurements at 6 mM SDS, the middle cluster is 7 mM SDS and the top cluster 8 mM. b) Data shown in a), normalized using Eq.~\ref{eq1}. The black line is a fit with a fixed slope of -2.}
\end{figure}

Including correction factors due to the concentration profile of the second droplet, as well as the autophoretic effect, we obtain the droplet velocity as a function of droplet-droplet separation:
\begin{equation}
\label{eq:U}
U(r) = M \frac{\partial G_s}{\partial G_e}\alpha \nabla c \approx \frac{2  K (c_{\infty}-c^*)(c_{cr}-c^*)al}{(2\eta_o+3\eta_i)(c_{cr}-c_{\infty})r^2}
\end{equation}
where we used the fact that $l/a\ll1$. The lengthscale $l=D/k$ is the only unknown parameter and all the others are fixed either by the experiment or obtained from the literature. Using this equation, we rescale all the data shown in Fig. ~\ref{fig4}a onto the mastercurve in Fig.~\ref{fig4}b. The log-log plot reveals a consistency with the predicted power law scaling with distance as $\propto 1/r^2$ (black line).

Using that $K=0.11$ mN m$^{-1} $ mM$^{-1}$ from the fit to the high SDS concentration regime of Fig. ~\ref{fig1}b, we find that $l=16\pm3$~nm, in agreement with the earlier observation that $l/a\ll1$ and that the dissolution process is diffusion limited. This length scale coincides with the size of an oily micelle, to within an order of magnitude. The obtained value for the fit parameter can be related to the critical Peclet number $\text{Pe}_{cr}$ at which self-propulsion occurs. When $l/a\ll1$, the critical Peclet number can be estimated by
\begin{equation}
 \label{eq:Pe}
 \text{Pe}= \frac{U(a) a}{\alpha D} \approx \frac{2  K  (c_\text{cr}-c^*)l}{(2\eta_o+3\eta_i)D}
\end{equation}
where we evaluated $U(a)/\alpha$ at $c_\infty= c_\text{cr}$.
Using the diffusion coefficient of oily micelles $D=10^{-10}$ m$^2$ s$^{-1}$~\cite{Clifford1966}, we find $\text{Pe}_{cr}=5$, which is in good agreement with the theoretically predicted value of $\text{Pe}_{cr}=4$ in~\cite{Michelin2013}. 

An interesting consequence of the fact that active droplets remove SDS from solution as they dissolve and move is that they leave behind them regions depleted of SDS that take time to equilibrate with their surroundings~\cite{Kranz2016}. As a result, we see memory effects, in which moving droplets can be repelled by the SDS-depleted trails of particles that had previously been at the same location. Supplementary Movie 3 shows an example of this type of memory effect.

These nonequilibrium interactions are relevant to both reactive and dissolving particles, while their strength depends on the rate at which the process occurs and the sensitivity of the particles to the surrounding gradient. The functional form is general for isotropic particles in a steady state and is expected to be universally applicable. 

\begin{acknowledgments}
We thank Mike Cates, Wilson Poon, and especially Eric Vanden-Eijnden for insightful discussions. We thank Gerhard Blab for providing the ray tracing simulations. This work was supported by the Materials Research Science and Engineering Center (MRSEC) program of the National Science Foundation under Award Number DMR-1420073 and by the NWO Graduate Program. E.W. and A.B. acknowledge financial support from the European Research Council under the European Union Seventh Framework Programme (FP/2007-2013)/ERC Grant Agreement no. [291667] HierarSACo. 
\end{acknowledgments}

\end {document}